\documentclass{PNASTWO}

\usepackage{graphicx}

\usepackage{PNASTWOF}

\usepackage{amssymb,amsfonts,amsmath}

\newcommand{\rr}{{\bf{r}}}
\newcommand{\kk}{{\bf{k}}}
\newcommand{\nn}{{\bf{n}}}
\newcommand{\pp}{{\bf{p}}}
\newcommand{\CC}{{\bf{C}}}
\DeclareMathOperator{\trace}{tr}
\newcommand{\rref}[1]{{\bf \ref{#1}}}
\contributor{Submitted to Proceedings
of the National Academy of Sciences of the United States of America}
\url{www.pnas.org/cgi/doi/10.1073/pnas.0709640104}
\copyrightyear{2008}
\issuedate{Issue Date}
\volume{Volume}
\issuenumber{Issue Number}

\begin{document}


\title{A favorably-scaling natural-orbital functional theory based on higher-order occupation probabilities} 





\author{Ralph Gebauer\affil{1}{ICTP -- The Abdus Salam International
  Centre for Theoretical Physics, Strada Costiera 11, 34151 Trieste,
  Italy}, Morrel H. Cohen\affil{2}{Department of Physics and
  Astronomy, Rutgers University, Piscataway, NJ
  08854}\affil{3}{Department of Chemistry, Princeton University,
  Princeton, NJ 08544}, \and Roberto
  Car\affil{3}{}\affil{4}{Department of Physics, Princeton University,
  Princeton, NJ 08544}}

\contributor{Submitted to Proceedings of the National Academy of Sciences
of the United States of America}

\significancetext{
Computations of the locations of the nuclei and the movement of
electrons within molecules and materials are widely used in science
and technology. Direct computation of a system's wave function for
that purpose becomes impractical as system size grows. Current
alternative methods can have difficulty with strongly-correlated
electron motion or spurious electron self-interaction. By using
``natural spin orbitals'' to describe the motion of individual electrons,
solving for them together with their joint and individual probabilities
of occurrence within the system, we are able to account better for
electron correlation when strong while avoiding self-interaction and
maintaining the growth of computation cost with system size at the
level of Hartree-Fock theory. Our numerical results for small test molecules
are excellent. 
}


\maketitle

\begin{article}

\begin{abstract}
We introduce a novel energy functional for ground-state
electronic-structure calculations. Its fundamental variables are the
natural spin-orbitals of the implied singlet many-body wave function
and their joint occupation probabilities. The functional derives from
a sequence of controlled approximations to the two-particle density
matrix. Algebraic scaling of computational cost with electron number
is obtainable in general, and Hartree-Fock scaling in the
seniority-zero version of the theory. Results obtained with the latter
version for saturated small molecular systems are compared with those
of highly-accurate quantum-chemical computations. The numerical
results are variational, capturing most of the correlation energy from
equilibrium to dissociation. Their accuracy is considerably greater
than that obtainable with current density-functional theory
approximations and with current functionals of the one-particle
density matrix only.
\end{abstract}






\section{Introduction}
\dropcap{C}omputing the ground-state energy of $N$ interacting electrons is
central to quantum chemistry, condensed-matter physics, and related
sciences. Reducing its complexity significantly below
that of the many-body wave function has been a major goal since
the early days of quantum mechanics. Density-functional theory (DFT)
\cite{Hohenberg1964, Kohn1965} achieved maximal reduction by using
electron density as the basic variable. DFT transformed many sciences and
technologies, but finding accurate, parameter-free approximations to
its exchange-correlation energy functional that avoid self-interaction
and capture strong electron correlation has remained difficult.
One-particle density-matrix  (1-DM) functional theories
\cite{Gilbert1975} have one
more degree of complexity. In them, the 1-DM is often
represented by its eigenvalues, the occupation numbers, and the
corresponding eigenvectors, the natural spin-orbitals (NSOs),
e.g.~\cite{Gritsenko2005, Lathiotakis2008}. While avoiding the
mean-field form of the 1-DM of DFT \cite{Kohn1965}, the 
approximations to the exchange-correlation functional of the 1-DM have
difficulties like those of the DFT approximations. Two-particle
density-matrix (2-DM) functional theories, e.g.~\cite{Garrod1964}, are
less reduced, an advantage. The ground-state energy is a
known, explicit functional of the 2-DM in Coulombic systems. However,
while necessary and sufficient conditions are known for the
$N$-representability of the 1-DM, no such conditions exist for the
2-DM~\cite{Coleman1963a}; reconstructing the $N$-particle wave function from the 2-DM is
a QMA-hard problem \cite{Liu2007}.  Nevertheless, major
progress has been 
made towards necessary conditions for $N$-representability that
can be systematically refined \cite{Zhao2004,Mazziotti2012}. While not
variational, the resulting calculations are almost as accurate as full
configuration interaction (FCI) calculations
\cite{Zhao2004,Nakata2001,Gidofalvi2008}. Their computational cost
scales as the 6th power of the basis-set size,
significantly worse than the asymptotic 3rd power scaling of
Hartree-Fock (HF) theory.
Here we introduce a new natural-orbital-functional theory, OP-NOFT, in
which the basic variables are the NSOs, their
occupation numbers, and their joint occupation probabilities (OP). The
latter allow us to represent the 2-DM accurately and transcend the
limitations of the 1-DM theories. Its general form contains
single-NSO through 4-NSO joint-occupation probabilities and scales
as the 5th power of the basis-set size. Its simplest formulation, for
seniority 0, OP-NOFT-0, corresponds to doubly-occupied
configuration interaction (DOCI) \cite{Weinhold1967}. It contains only
single- and 2-natural-orbital (NO) OPs and retains the 3rd power
scaling of HF energy-functional minimization, albeit with a higher
prefactor. 
It describes the dissociation of simple diatomic molecules and
multi-atom chains with accuracy comparable to that of DOCI, which
uses a compact basis of Slater determinants (SD) but retains
exponential scaling. OP-NOFT-0 is powerful at high correlation,
i.e.~for static correlation at intermediate and large interatomic
separations where HF fails due to the multi-reference character of the
ground-state wavefunction. There, OP-NOFT-0 outperforms HF, DFT and
quantum-chemistry methods such as (single-reference) coupled cluster
with single, double and perturbative triple electron-hole excitations
(CCSD(T)), a standard of accuracy near equilibrium separations. This
introduction of higher-order OPs as variational parameters, with
closure of the theory at their level, is the essential novelty of our
work and is responsible for its favorable scaling with high accuracy.

\section{OP-NOFT, general formulation}
\subsection{The NSO basis}

We consider time-reversal invariant saturated systems
with non-degenerate, singlet ground states. The inverse approach
\cite{Garrod1964} starts from a set of $N$-representability conditions
on the 2-DM needed for it to be derivable from a generic
$N$-electron wavefunction. Instead, we take a forward approach: we
introduce a specific form for the trial wavefunction and derive
the 2-DM explicitly. Our starting point
is that of conventional FCI, except that our
one-particle basis is the complete set of NSOs 
of the trial function $\Psi$,
$\psi_k(x) = \phi_k(\rr) \chi_k(\sigma)$, 
with $\rr$ space and $\sigma$ spin
coordinates. The NOs
$\phi_k(\rr)$ are real and independent of the spin function
$\chi_k(\sigma)$. The complete set of $N$-electron
orthonormal SDs $\Phi_{\kk}(x_1,x_2,\cdots,x_N)$, 
$\kk=k_1,k_2,\cdots,k_N$, formed from its NSOs 
supports representation of any trial wavefunction
$\Psi(x_1,x_2,\cdots,x_N)$ as the expansion
\begin{equation}
\Psi(x_1,\cdots,x_N) = \sum_{\kk} C_{\kk} \Phi_{\kk}(x_1,\cdots,x_N).
\label{expansion}
\end{equation}

As the ground-state wave function is real, so are the trial functions
and the normalized $C_{\kk}$  ($\sum_{\kk} C_{\kk}^2 = 1$).
As the trial function or the coefficients vary in the search for the
ground state, so do the NSOs, as in any NOFT. The exponential
complexity of determining the ground-state energy by variation of the
$C_{\kk}$ is composed of the separate exponential complexities of the
signs and the magnitudes of the coefficients $C_{\kk}$. We use
distinct reductive approximations for their signs and 
magnitudes. The signs and consequently the sign-approximation scheme
depend on the sign convention chosen for the SDs. We use the Leibniz
form for the SDs,
\begin{equation}
\Phi_{\kk}(x_1,\cdots,x_N) =
\frac1{\sqrt{N!}} \sum_p \text{sgn}\{P_p\} P_p \psi_{k_1}(x_1) \cdots
\psi_{k_N}(x_N).
\label{SDs}
\end{equation}
The sum is over the elements of the symmetric group of order $N$, the
permutations $P_p$. The sign of $\Phi_{\kk}$ is fixed by the ordering
$k_1 < k_2 < \cdots < k_N$ in the product of the NSO $\psi_{k_i}$ in
\rref{SDs}. Once this sign convention is understood, the SDs and their
coefficients can be specified simply by listing the NSOs occupied
in the SDs, i.e.~by the index $\kk$.

\subsection{The 1-DM, the orthogonality constraint, the PDC}

The 1-DM of $\Psi$,
\begin{eqnarray}
\rho(x',x) = N \int dx_2\cdots dx_N \nonumber \\
\Psi(x',x_2,\cdots,x_N) \Psi^*(x,x_2,\cdots,x_N),
\label{rho}
\end{eqnarray}
becomes
\begin{equation}
\rho(x',x) = \left[ \sum_{\kk; i,j\not\in\kk} C_{i,\kk} C_{j,\kk}
  \right] \, \psi_i(x') \psi_j(x)
\label{rho1_spin}
\end{equation}
after \rref{expansion} and \rref{SDs} are inserted into
\rref{rho}. In \rref{rho1_spin} the subindex $\kk$ specifies 
the $N-1$ NSOs present in $\Phi_{i,\kk}$ and $\Phi_{j,\kk}$, 
excluding $\psi_i$ and $\psi_j$. As the $\psi$ are
the NSOs of $\Psi$, the eigenfunctions of $\rho(x',x)$,
the bracketed quantity in \rref{rho1_spin} must vanish for $i\neq
j$. Regard the coefficients $C_{i,\kk}$ and $C_{j,\kk}$
as the components of vectors $\CC_i$ and $\CC_j$
and the bracket as their scalar product $\CC_i \cdot \CC_j$,
which must vanish. There are two realizations of this
orthogonality constraint, a condition of consistency between $\Psi$
and the $\psi$. In the first, an {\em inclusive} and most general
form (OC), the presence of $\Phi_{i,\kk}$ in $\Psi$ does not exclude
the presence of $\Phi_{j,\kk}$. The individual terms in the
scalar product need not vanish, only their sum must. In the second, an
{\em exclusive} form, the pair-difference constraint (PDC) is
a special case of the OC, in which the presence of $\Phi_{i,\kk}$
excludes $\Phi_{j,\kk}$ so that either 
$C_{i,\kk}$ or $C_{j,\kk}$ is zero for each $\kk$, and the sum
vanishes term by term. Under the PDC, those $\Phi$ present
in the expansion of $\Psi$ must differ from one another by at least
two NSOs, forming a subspace 
$\{\Phi\}_{\text{PDC}}$ of the SD space $\{\Phi\}$ and restricting the trial
function space $\{\Psi\}$ to $\{\Psi\}_{\text{PDC}}$. The OC is a
necessary and sufficient condition for $N$-representability, whereas
the PDC is only sufficient. We impose the PDC on the $\Phi$ as a
simplifying variational approximation. 

Under the OC or PDC, $\rho(x',x)$ takes the diagonal
form 
\begin{equation}
\rho(x',x) = \sum_k p_1(k) \, \psi_k(x')\psi_k(x).
\label{rho1diag}
\end{equation}
Here, the $p_1(k) = \sum_{\nn} C_{\nn}^2 \, \nu_{k,\nn}$, where $\nu_{k,\nn}
= 1$ if $k \in \nn$ and $0$ otherwise, are the eigenvalues of
$\rho(x',x)$, the occupation numbers or occupation probabilities
(1-OP) of its eigenfunctions $\psi_k$. They satisfy the necessary and
sufficient conditions $0 \le p_1(k) \le 1$ and $\sum_k p_1(k) = N$. In
general, only $M > N$ occupation numbers $p_1(k)$ are non-negligible,
and only the corresponding active NSOs need be included
in the representation of any trial function, providing a natural
cutoff. The 1-DM is thus of algebraic complexity in the
1-OPs and the NSOs, as $M$ scales linearly with $N$. 

\subsection{The 2-DM, the sign conjecture, the $\xi$-approximation}

The 2-DM of $\Psi$,
\begin{eqnarray*}
\pi(x'_1x'_2;x_1x_2) = N (N-1) \int dx_3\cdots dx_N \\
\Psi(x'_1,x'_2,x_3,\cdots,x_N) \Psi^*(x_1,x_2,x_3,\cdots,x_N),
\end{eqnarray*}
becomes
\begin{eqnarray*}
\pi(x'_1x'_2;x_1x_2) &=& 
\sum_{\scriptscriptstyle
\overset{i<i', j<j',\kk}
{i,i',j,j' \not\in \kk}}
C_{ii'\kk} C_{jj'\kk}\\
&&\left(
\psi_i(x'_1)\psi_{i'}(x'_2)-\psi_{i'}(x'_1)\psi_i(x'_2)\right)\\
&&\left(
\psi_j(x_1)\psi_{j'}(x_2)-\psi_{j'}(x_1)\psi_j(x_2)\right).
\end{eqnarray*}
$\pi$ separates into a part $\pi^d$ diagonal in the indices, i.e.~with
$ii' = jj'$, and an off-diagonal part, 
$\pi^{od}$, with $ii' \neq jj'$:
\begin{eqnarray}
\pi^d(x'_1x'_2;x_1x_2) &=& \frac1{2}\sum_{i\ne j} p_{11}(ij) \nonumber\\
&&\left(
\psi_i(x'_1)\psi_j(x'_2)-\psi_j(x'_1)\psi_i(x'_2)\right)\nonumber\\
&&\left(
\psi_i(x_1)\psi_j(x_2)-\psi_j(x_1)\psi_i(x_2)\right)
\label{pi-d}\\
\pi^{od}(x'_1x'_2;x_1x_2) &=& 
\sum_{\scriptscriptstyle
\overset{i<i' \ne j<j',\kk}
{i,i',j,j' \not\in \kk}}
C_{ii'\kk} C_{jj'\kk}\nonumber\\
&&\left(
\psi_i(x'_1)\psi_{i'}(x'_2)-\psi_{i'}(x'_1)\psi_i(x'_2)\right)\nonumber\\
&&\left(
\psi_j(x_1)\psi_{j'}(x_2)-\psi_{j'}(x_1)\psi_j(x_2)\right)
\label{pi-od}
\end{eqnarray}

Electron correlation is expressed through
$\pi^{od}$. The analogous off-diagonal part of
$\rho(x',x)$ is suppressed by the OC, an advantage of the NSO
basis. Note also that the PDC has eliminated 3-index terms from
$\pi^{od}$ in~\rref{pi-od}. 

The $p_{11}(ij) = \sum_{\nn} C_{\nn}^2 \, \nu_{i,\nn} \nu_{j,\nn}$ in
$\pi^d$ are joint 2-state occupation probabilities (2-OPs).
Establishing the $N$-representability conditions for the 2-OPs is a hard
problem because new conditions constrain the 2-OPs that arise from the
positivity conditions for the $q$-OPs, i.e.~the $p_{11\cdots 1}
(i_1,i_2,\cdots,i_q) = \sum_{\nn} C_{\nn}^2 \, \nu_{i_1,\nn}
\nu_{i_2,\nn} \cdots \nu_{i_q,\nn}$, at any
order $q$. 
These conditions derive from the so-called $(2,q)$ positivity
conditions \cite{Mazziotti2012a} 
restricted to the diagonal elements of the 2-DM \cite{Ayers2007}.
Limiting ourselves to the $(2,2)$ and $(2,3)$
conditions, the following conditions for the 2-OPs hold:
\begin{eqnarray}
\mbox{sup}(p_1(i)+p_1(j)-1,0) &\le& p_{11}(ij) \,\le\, p_1(<) \label{bound1}\\
\mbox{sup}(p_1(i)+p_1(j)+p_1(k)-1,0) &\le&
p_{11}(ij)+\nonumber\\
&&p_{11}(ik)+p_{11}(jk) \label{bound2}
\end{eqnarray}
$p_1(<)$ is the lesser of $p_1(i)$ and $p_1(j)$. In addition the sum rule
\begin{eqnarray}
\sum_{j (\ne i)} p_{11}(ij) &=& (N-1) p_1(i) \label{p11sum}
\end{eqnarray}
must be satisfied. Conditions \rref{bound1}--\rref{p11sum} were first
established by Weinhold and Bright Wilson \cite{Weinhold1967a}. 
They are necessary but not sufficient conditions for
$N$-representability \cite{Ayers2007,Davidson1969}. Establishing a
complete set of conditions is a QMA-hard problem because the number of
$(2,q)$ positivity conditions increases combinatorially with
increasing $q$. 
Fortunately numerical calculations on atoms and molecules indicate
that sufficiently accurate lower-bound ground-state energies often
result by imposing $(2,q)$-positivity conditions with $q\le 3$
\cite{Zhao2004,Mazziotti2012b}. This
suggests that even in the most difficult situations, fermionic problems
in atoms and molecules should require only a finite and small set of positivity
conditions. Here we shall limit ourselves to conditions
\rref{bound1}--\rref{p11sum}, as we found in our numerical
calculations that they are sufficient to produce accurate
lower-bounds. If higher order conditions were found to be necessary,
it would not be hard to add a few more of them in the numerical scheme
presented later.  

The $\pi^d$ of \rref{pi-d} contains only 2-OPs and products of 2
distinct NSOs; it has at most algebraic complexity $\sim M^3$ deriving
from condition \rref{bound2}. Thus when only conditions
\rref{bound1}--\rref{p11sum} are imposed, the exponential
complexity of the ground-state problem resides entirely in the
$\pi^{od}$ of \rref{pi-od}. We extract
the sign $s(ii'\kk)$ of the coefficient $C_{ii'\kk}$ in
\rref{pi-od} and, relating its magnitude to the joint $N$-OP
$p_{11\cdots 1}(ii'\kk) \equiv C_{ii'\kk}^2$,
we rewrite~\rref{pi-od} as
\begin{align}
\pi^{od}(x'_1x'_2;x_1x_2) =& 
\sum_{\scriptscriptstyle
\overset{i<i'\ne j<j',\kk}
{i,i',j,j' \not\in \kk}}
s(ii'\kk) s(jj'\kk) \nonumber \\
&p^{1/2}_{11\cdots 1}(ii'\kk) 
p^{1/2}_{11\cdots 1}(jj'\kk)\nonumber\\
&\left(
\psi_i(x'_1)\psi_{i'}(x'_2)-\psi_{i'}(x'_1)\psi_i(x'_2)\right)\nonumber\\
&\left(
\psi_j(x_1)\psi_{j'}(x_2)-\psi_{j'}(x_1)\psi_j(x_2)\right).
\label{pi-od-signs}
\end{align}
We suppose that a variational approximation exists in which
\begin{equation}
s(ii'\kk) s(jj'\kk) = s(ii') s(jj'), \forall\kk.
\label{signconjecture}
\end{equation}
The {\em sign conjecture} \rref{signconjecture} reduces the sign
complexity to algebraic, scaling as $M^2$. $\pi^{od}$ 
simplifies to
\begin{eqnarray}
\pi^{od}(x'_1x'_2;x_1x_2) &=& 
\sum_{\scriptscriptstyle
{i<i' \ne j<j'}}
s(ii') s(jj') \nonumber \\
&&\left[ \sum_{\scriptscriptstyle \overset{\kk}
{i,i',j,j' \not\in \kk}}
p^{1/2}_{11\cdots 1}(ii'\kk) 
p^{1/2}_{11\cdots 1}(jj'\kk) \right] \nonumber\\
&&\left(
\psi_i(x'_1)\psi_{i'}(x'_2)-\psi_{i'}(x'_1)\psi_i(x'_2)\right)\nonumber\\
&&\left(
\psi_j(x_1)\psi_{j'}(x_2)-\psi_{j'}(x_1)\psi_j(x_2)\right).
\label{pi-od-signs1}
\end{eqnarray}
The quantities $p^{1/2}_{11\cdots 1}(ii'\kk)$ and $p^{1/2}_{11\cdots
  1}(jj'\kk)$ are $\kk$-th components of vectors
$\pp^{1/2}_{11\cdots 1}(ii')$ and $\pp^{1/2}_{11\cdots 1}(jj')$. The
bracketed quantity in~\rref{pi-od-signs1} is their scalar
product. Express it as
\begin{eqnarray}
\sum_{\scriptscriptstyle \overset{\kk}
{i,i',j,j' \not\in \kk}}
p^{1/2}_{11\cdots 1}(ii'\kk) 
p^{1/2}_{11\cdots 1}(jj'\kk) &=& \nonumber \\
p^{1/2}_{1100}(ii'jj') p^{1/2}_{0011}(ii'jj') \xi(ii'jj'),
\label{scalar}
\end{eqnarray}
where $p_{1100}(ii'jj')$ is the square magnitude of the vector
$\pp^{1/2}_{11\cdots 1}(ii')$ and $p_{0011}(ii'jj')$ that of
$\pp^{1/2}_{11\cdots 1}(jj')$. $p_{1100}(ii'jj')$ is the probability
that $\psi_i$ and $\psi_{i'}$ are occupied while $\psi_j$ and
$\psi_{j'}$ are not:
\begin{eqnarray*}
p_{1100}(ii'jj') &=& \sum_{\scriptscriptstyle \overset{\kk}
{i,i',j,j' \not\in \kk}} p_{11\cdots 1}(ii'\kk) \\
&=& \sum_{\nn} C_{\nn}^2 \, \nu_{i,\nn} \nu_{i',\nn} \, 
\left( 1 - \nu_{j,\nn} \right) \left( 1 - \nu_{j',\nn} \right),
\end{eqnarray*}
and the reverse is true for $p_{0011}(ii'jj')$. 

The Schwarz inequality $0 \le \xi(ii'jj') \le 1$ imposes bounds on
$\xi(ii'jj')$, the cosine of the hyper-angle between the
vectors. Substituting~\rref{scalar} into~\rref{pi-od-signs1} yields
\begin{eqnarray}
\pi^{od}(x'_1x'_2;x_1x_2) &=& 
\sum_{\scriptscriptstyle
i<i' \ne j<j'}
s(ii') s(jj') \nonumber\\
&&\left[p_{1100}(ii'jj') \,p_{0011}(ii'jj')\right]^{1/2}
\xi(ii'jj') \nonumber\\
&&\left(
\psi_i(x'_1)\psi_{i'}(x'_2)-\psi_{i'}(x'_1)\psi_i(x'_2)\right)\nonumber\\
&&\left(
\psi_j(x_1)\psi_{j'}(x_2)-\psi_{j'}(x_1)\psi_j(x_2)\right),
\label{final_pi-od}
\end{eqnarray}
in which only $\xi(ii'jj')$ retains exponential complexity:
\begin{displaymath}
\xi(ii'jj') = \frac{
\sum' p^{1/2}_{11\cdots 1}(ii'\kk) p^{1/2}_{11\cdots 1}(jj'\kk)}
{\left(
\sum' p_{11\cdots 1}(ii'\kk) \, \,
\sum' p_{11\cdots 1}(jj'\kk) 
\right)^{1/2}},
\end{displaymath}
where the primed sums are over all $\kk$ with $i,i',j,j' \not\in \kk$.

Inserting 4-OPs like
\begin{displaymath}
p_{1111}(ii'kl) = \sum_{\nn} C_{\nn}^2 \, \nu_{ii',\nn} \,
\nu_{kl,\nn}; \qquad k < l \neq i,i',j,j'
\end{displaymath}
in place of the $N$-OPs in $\xi$ reduces the complexity of $\pi^{od}$ to
algebraic. The resulting approximation,
\begin{equation}
\xi(ii'jj') \approx \frac{
\sum''_{k<l} p^{1/2}_{1111}(ii'kl) p^{1/2}_{1111}(jj'kl)}
{\left(
\sum''_{k<l} p_{1111}(ii'kl) \, \,
\sum''_{k<l} p_{1111}(jj'kl) 
\right)^{1/2}},
\label{xi-4}
\end{equation}
is not variational, but obeys the 0,1 bounds of the Schwarz
inequality. It is exact for $N=4$ within the PDC, and scales as
$M^4$. In~\rref{xi-4} the doubly primed sums are over the indices
$k<l$, which must differ from $i,i',j,j'$.
Bounds on the $p_{1111}$ that are the
generalizations of \rref{bound1}--\rref{p11sum} for 3-OPs and 4-OPs
can be formulated. 

\subsection{The OP-NOFT energy functional}

\ The trial energy, $E[\Psi] = \left( \Psi, \hat{H}\Psi \right)$, the
expectation value of the Hamiltonian $\hat{H}$, is an
explicit functional of the 1- and 2-DM:
\begin{displaymath}
E[\Psi] = E[\rho, \pi] = \trace\left\{\rho \hat{h}\right\} +
\trace\left\{ \pi \hat{w} \right\}.
\end{displaymath}
Here $\hat{h}$ is the single-particle kinetic-energy operator plus the
external potential, and $\hat{w}$ is the 2-electron Coulomb
interaction. $E[\rho, \pi]$ splits into two parts, $E^d$
diagonal and $E^{od}$ off-diagonal in the SD:
\begin{eqnarray}
E &=& E^d + E^{od} \nonumber \\
E^d &=& \trace\left\{\rho \hat{h}\right\} + \trace\left\{\pi^{d}
\hat{w}\right\} \nonumber \\
E^{od} &=& \trace\left\{\pi^{od} \hat{w}\right\}.
\label{Esum}
\end{eqnarray}
The HF wave function minimizes $E^d$; $\pi^{od}$ introduces electron
correlation into $E^{od}$. The explicit forms of $E^d$ and $E^{od}$
follow from \rref{rho1diag}, \rref{pi-d}, and \rref{final_pi-od}:
\begin{equation}
E^d = \sum_i p_1(i) h_{ii} + \sum_{i<j} p_{11}(ij) \left[ {\cal
    J}_{ij} - {\cal K}_{ij}\right],
\label{E-d}
\end{equation}
where $h_{ii} = (\psi_i, \hat{h}\psi_i)$, and  ${\cal J}_{ij} = 
(\psi_i\psi_i,\hat{w}\psi_j\psi_j)$ and ${\cal K}_{ij} = 
(\psi_i\psi_j,\hat{w}\psi_j\psi_i)$ are the usual Coulomb and exchange 
integrals, respectively.
\begin{eqnarray}
E^{od} &=& \sum_{\scriptscriptstyle i<i' \ne j<j'} s(ii') s(jj')
p^{1/2}_{1100}(ii'jj') p^{1/2}_{0011}(ii'jj')
\nonumber \\
&& \xi(ii'jj') \left[ {\cal K}_{ii'jj'} - {\cal K}_{ii'j'j} \right],
\label{E-od}
\end{eqnarray}
where ${\cal K}_{ii',jj'} = (\psi_i\psi_{i'},\hat{w}\psi_j\psi_{j'})$.
\rref{Esum} -- \rref{E-od} define the OP-NOFT energy
functional. Including the complexity of efficient
evaluation of the matrix elements, it scales as $M^5$ if the
$N$-representability conditions for the 3- and 4-OPs can be limited to
those deriving from the $(3,q)$ and $(4,q)$ positivity conditions with $q
\le 4$.  

\subsection{Proof of the sign conjecture}

A variational sign approximation must be a statement
about the sign $s(ii'\kk)$ or $s(jj'\kk)$  of each coefficient appearing
in~\rref{pi-od}. To prove the sign
conjecture \rref{signconjecture}, we must find one in which the
$\kk$-dependences of $s(ii'\kk)$ and  $s(jj'\kk)$ cancel out. 
One, for the general case of matrix elements 
$\left[ {\cal K}_{ii',jj'} - {\cal K}_{ii',j'j} \right]$
of arbitrary sign, is presented here. Another, valid only for positive
matrix elements, is presented in Section S1 of the Supporting
Information (SI). The two approximations yield the same results for
OP-NOFT-0. 

{\em Arbitrary signs of matrix elements:}
Assigning each index $l$ in $C_{\kk}$ a sign $s(l)$ and taking
$s(\kk)$ as their product,
\begin{displaymath}
s(\kk) = \prod_{l \in \kk} s(l),
\end{displaymath}
is a variational approximation. Consequently $s(ii'\kk) = s(i) s(i')
s(\kk)$ and
\begin{equation}
s(ii'\kk) s(jj'\kk) = s(i) s(i') s(j) s(j'),
\label{signprod}
\end{equation}
and the sign conjecture is proved for arbitrary signs of matrix
elements. This approximation treats the form and phase, $0$ or $\pi$,
of each NSO as independent variables. The choice of signs for each
index is not specified in \rref{signprod}. Most computation
schemes start their convergence towards the minimum energy with random
initial NSOs, and similarly the choice of signs in \rref{signprod}
should be random, half positive and half negative. The number $L$ of
initial NSOs should be greater than $M$ to allow for the possibility of
unequal numbers of positive and negative signs associated with the $M$
active NSOs.

\section{OP-NOFT-0}

The SD's in \rref{expansion} can be classified by their seniority,
the number $A$ of singly-occupied one-particle states they
contain. For $N$ even and for a global
spin singlet ($S=0$) state, the $N$-particle Hilbert space divides into
sectors of increasing even seniority starting with $A=0$, where all
SD's contain only doubly occupied states. 
That only even seniority occurs is a consequence of the orthogonality constraint. 
For molecular systems
CI expansions converge rapidly
with seniority, and $A=0$ calculations (DOCI calculations) describe dissociation rather well, as
demonstrated in~\cite{Bytautas2011}. 

We now formulate OP-NOFT explicitly in the $A=0$ sector both to
illustrate further how an OP-NOFT functional is constructed and to prepare  
for numerical implementation; it becomes OP-NOFT-0, in which the PDC
is automatically satisfied. Tracing out the spins,
\rref{rho1diag} becomes:
\begin{equation}
\rho(\rr',\rr) = 2 \sum_k p_1(k) \, \phi_k(\rr') \phi_k(\rr).
\label{rho1}
\end{equation}
$k$ now labels $M(>N/2)$ active doubly-occupied NO states, and the following conditions
hold:
\begin{equation}
0 \le p_1(k) \le 1 \mbox{ and } 2\sum_k p_1(k) = N.
\label{rho1cond}
\end{equation}

In \rref{rho1} and \rref{rho1cond} $p_1(k)$ is the occupation number of either 
of the paired NSOs having the NO $\phi_k$.

In the 2-DM, the impact of double occupancy on the structure of $\pi^d$ is 
minor, but it results in a major simplification of the structure of 
$\pi^{od}$. We make the orbital and spin components of the NSO indices explicit, so
that they take the form $is$, with $i$ now the orbital index and $s = \pm$ the
spin index. Because of double occupancy, the only index pairs that can enter
$\pi^{od}$ in \rref{final_pi-od} are $i+,i-$ and $j+,j-$. Similarly, the
only sets of two index pairs that can enter the rhs of \rref{xi-4} are $i+,i-$,
$k+,k-$ and $j+,j-$, $k+,k-$. The occupation numbers $\nu_{i+}$ and 
$\nu_{i-}$ are always equal, with values 0 or 1, so that all 4-NSO OPs in
\rref{xi-4} and \rref{final_pi-od} are identical to the corresponding 
spin independent 2-NSO OPs, e.g.~$p_{1111}(i+i- k+k-) = p_{11}(ik)$.
Correspondingly, the signs in \rref{final_pi-od} depend only on a single
orbital index, $s(i+i-) = s(i)$, and the $\xi$ depend only on two-orbital
indices, $\xi(i+i- j+j-) = \xi(ij)$. With these simplifications, the 2-DM of
\rref{pi-d} and \rref{final_pi-od} becomes
\begin{equation}
\pi(\rr'_1 \rr'_2;\rr_1 \rr_2) =
\pi^d(\rr'_1 \rr'_2;\rr_1 \rr_2) +
\pi^{od}(\rr'_1 \rr'_2;\rr_1 \rr_2),
\label{pisum}
\end{equation}
 after tracing out the spins, with
\begin{eqnarray}
\pi^d(\rr'_1 \rr'_2;\rr_1 \rr_2) &=& 
2 \sum_{ij} p_{11}(ij) \nonumber\\
&&\big(2 \phi_i(\rr'_1) \phi_j(\rr'_2) \phi_i(\rr_1) \phi_j(\rr_2) - \nonumber \\
&&\phi_i(\rr'_1) \phi_j(\rr'_2) \phi_j(\rr_1) \phi_i(\rr_2) \big) 
\label{pi-d_DOCI}\\
\pi^{od}(\rr'_1 \rr'_2;\rr_1 \rr_2) &=& 2 \sum_{i\ne j} s(i) s(j) \nonumber \\
&&\left[ p_{10}(ij) p_{01}(ij)\right]^{1/2} \xi(ij) \nonumber \\
&& \phi_i(\rr'_1) \phi_i(\rr'_2) \phi_j(\rr_1) \phi_j(\rr_2).
\label{pi-od_DOCI}
\end{eqnarray}
The sum in \rref{pi-d_DOCI} includes the term $i=j$, for which $p_{11}(ii) =
p_1(i)$ because of double occupancy, and $\xi(ij)$ in \rref{pi-od_DOCI} is now
\begin{equation}
\xi(ij) \approx \frac{
\sum_{k(\ne i,j)} p^{1/2}_{11}(ik) p^{1/2}_{11}(jk)
}{\left[
\sum_{k(\ne i,j)} p_{11}(ik) \, \, \sum_{k(\ne i,j)} p_{11}(jk)
\right]^{1/2}}.
\label{xi_DOCI}
\end{equation}

The one- and two-orbital OPs of OP-NOFT-0 lie within the same bounds as 
in the general case, \rref{bound1}--\rref{bound2}, 
if we impose only the $(2,2)$ and $(2,3)$ positivity conditions, 
and their sum rules
become, respectively, \rref{rho1cond} and
\begin{equation}
2 \sum_{j(\ne i)} p_{11}(ij) = (N-2) p_1(i).
\label{p11sum_DOCI}
\end{equation}

The $\pi$ of \rref{pisum} satisfies two important sum rules
\begin{eqnarray*}
\int d\rr_2 \, \, \pi(\rr \rr_2; \rr' \rr_2) &=& (N-1) \rho(\rr,\rr') \\
\int d\rr_1 \, d\rr_2 \, \, \pi(\rr_1 \rr_2;\rr_1 \rr_2) w(r_{12})
&\ge& 0. 
\end{eqnarray*}

The OP-NOFT-0 form for $\pi$, \rref{pisum}--\rref{xi_DOCI}, 
is exact 
when $N=2$ with $\xi=1$. It is equivalent to DOCI for $N=4$. When $N>4$, the 
$\xi$-approximation of \rref{xi_DOCI} 
and the assumption on the $N$-representability condition
break the equivalence to DOCI. Our
numerical results presented below and our detailed examinations of its 
formal structure at dissociation suggest that it is a very good approximation
to DOCI, with algebraic instead of exponential complexity.

The expectation value $E = \langle \Psi | H |
\Psi \rangle$ of the Hamiltonian $H = h + \frac1{2} \sum_{i\ne j}
w(r_{ij})$, where $h$ is a sum of one-body terms, becomes: 
\begin{eqnarray}
E &=& 2 \sum_i p_1(i) \langle\phi_i | h | \phi_i\rangle + \nonumber \\
&&\sum_{ij} p_{11}(ij) \left(2 J_{ij} - K_{ij}\right) + \nonumber \\
&&\sum_{i\ne j} s(i) s(j) p_{10}^{1/2}(ij)\,p_{01}^{1/2}(ij)\,
\xi(ij) \, K_{ij}, \label{energy}
\end{eqnarray}
where $J_{ij}$ and $K_{ij}$ are Hartree- and exchange integrals
defined in terms of the NOs $\{\phi_i\}$.
With \rref{xi_DOCI} for $\xi(ij)$, $E$
in \rref{energy} is a functional of the NOs, 
and the 1- and 2-state OPs. 
The signs are chosen a priori by a sign rule and are not variables.
$p_{10}$ is
related to $p_{11}$  and $p_1$  by $p_1(i) = p_{11}(ij) +
p_{10}(ij)$ and can be eliminated from the functional.
Each sum in the denominator of $\xi(ij)$ in \rref{xi_DOCI} can be simplified 
by use of the sum rule of \rref{p11sum_DOCI} to, e.g.,
\begin{displaymath}
\sum_{k(\ne i,j)} p_{11}(ik) = \frac1{2} (N-2) p_1(i) - p_{11}(ij).
\end{displaymath}

As stated in the general section on the 2-DM, we assume that the $(2,2)$
and $(2,3)$ positivity conditions are sufficient in practice. Were this
not the case, it would be straightforward to add a few more positivity
conditions to achieve practical sufficiency. Under this circumstance,
the infimum of $E$  with respect to the
NOs and the OPs, subject to the constraints \rref{rho1cond} and
\rref{bound1}, \rref{bound2}, \rref{p11sum_DOCI}, yields a variational
approximation to the ground-state energy, 
apart from the $\xi$-approximation for $N>4$.
 
The last term on the
rhs of \rref{energy} 
originates from $\pi^{od}$  and drives electron correlation; without it 
the infimum is the HF energy. The second
term on the rhs of \rref{energy} originates from
$\pi^d$, the form of which is represented exactly in our theory. It contains only
positive contributions and is essential; the integral
relation connecting $\pi$ and $\rho$ depends only on $\pi^d$ and
guarantees that the functional $E$
is self-interaction free. \rref{energy} is a generalization
of the NOFT formulations of 1-DM functional theories, which 
require only 1-state OPs. The extra complexity from
2-state OPs and implicit 4-state OPs is more than compensated by the substantial gain in
accuracy it makes possible. The computational cost of calculating $E$ 
from \rref{energy} scales like HF
energy-functional minimization with a greater prefactor ($M^3$ vs $(N/2)^3$)
due to fractional occupation of NOs. 

\section{Numerical results for simple molecular systems}

To test this $A=0$ version, we studied several diatomic molecules and
linear chains of H atoms with open boundary conditions.  We included
all electrons (core and valence) and expanded the NOs in a contracted
Gaussian 6-31G$^{**}$ basis, unless otherwise specified. The
constrained minimization of the functional was performed by damped
Car-Parrinello dynamics \cite{Car1985}, as detailed in Section S2 of
the SI.
 
We started the minimization from NOs and OPs obeying the constraints
but otherwise random.  The matrix element $K_{ij}$ is positive in
\rref{energy}.  The signs were therefore taken from the sign rule of
the Table S1 in Sec.~S1 of the SI, except that the $(1,1)$ case does
not occur in zero seniority: for $i \le N/2$, $R(i+,i-)=0$, and $s(i)=
+1$; for $i > N/2$, $R(i+,i-)=2$, and $s(i)= -1$. They were kept fixed
during optimization.

At convergence, a subset of NOs, the active NOs, had $p_1 \ge
10^{-3}$. The remaining NOs contributed negligibly to the energy. The
same active NOs and signs were also found for several test cases
starting from a sufficiently large set of random NOs, half with
positive and half with negative signs\footnote{For H$_2$ the sign rule
  ($s(i\le N)=+1$ and $s(i>N)=-1$) holds near equilibrium, but a more
  complex pattern emerges at large separation where additional
  positive signs are needed for the van der Waals tail of the
  interaction potential \cite{Sheng2013}. In principle, these positive
  signs could be obtained with our minimization procedure, but their
  effect is beyond the accuracy of the present calculations}.  It is
significant for the rule of \rref{signprod} for arbitrary
matrix-element signs that for the systems tested, the PWBT-based rule
of the Table and the alternative of random initial assignment of signs
to pairs yield the same results for positive matrix elements. The
procedure of \rref{signprod} also yields half positive and half
negative signs for the pairs when signs are assigned randomly to the
individual NSOs with no reference to the matrix-element signs.

We also performed restricted HF, DFT (PBE \cite{Perdew1996} and/or
PBE0 \cite{Adamo1999}), and CCSD(T) calculations, with the same
basis. The relatively small 6-31G$^{**}$ set is adequate for the
comparisons of interest.  We report the dissociation energy curves of
the dimers H$_2$, LiH and HF in Figs. S1, S2, and S3 in Section S3 of
the SI. For H$_2$, our functional depends only on 1-state OPs and
reduces to the exact expression of Löwdin and Shull
\cite{Lowdin1956}. The OP-NOFT-0 dissociation energy curve thus
coincides with CASSCF at all interatomic separations.  Even in a
system as simple as H$_2$, (spin-restricted) HF and DFT fail badly at
dissociation because these single reference theories cannot recover
the Heitler-London form of the wavefunction.  The 4-electron case of
LiH provides the first test of 2-state OPs.  The conditions
\rref{bound1},\rref{bound2}, and \rref{p11sum_DOCI} simplify in this
case as discussed in Section S4 of the SI. Expression \rref{xi_DOCI}
for $\xi$ is exact here, but the restriction to $A=0$ is not. That the
OP-NOFT-0 dissociation curve of LiH almost coincides with CASSCF
indicates that higher seniorities contribute negligibly to its
ground-state energy. Finally, HF, a 10-electron system, provides the
first complete test of the theory. Here the OP-NOFT-0 dissociation
energy curve follows that of CASSCF with a positive energy shift over
the entire separation range. This indicates that the $\xi$ and $A=0$
approximations work well. We cannot exclude compensation of errors
between the two approximations, but the results below suggest that it
is not the main factor behind their quality.

The results in Section S3 of the SI depict the breaking of a
single-bond in simple molecules. To assess the performance of
OP-NOFT-0 in more challenging situations with many bonds or a
multiple-bond, we studied the symmetric dissociation of linear H
chains and the dissociation of the N$_2$ dimer. These cases have been
used to test compact CI expansions controlled by seniority
\cite{Bytautas2011}; FCI calculations are thus available for
comparison.
\begin{figure}
\begin{center}
  \includegraphics[width= \columnwidth]{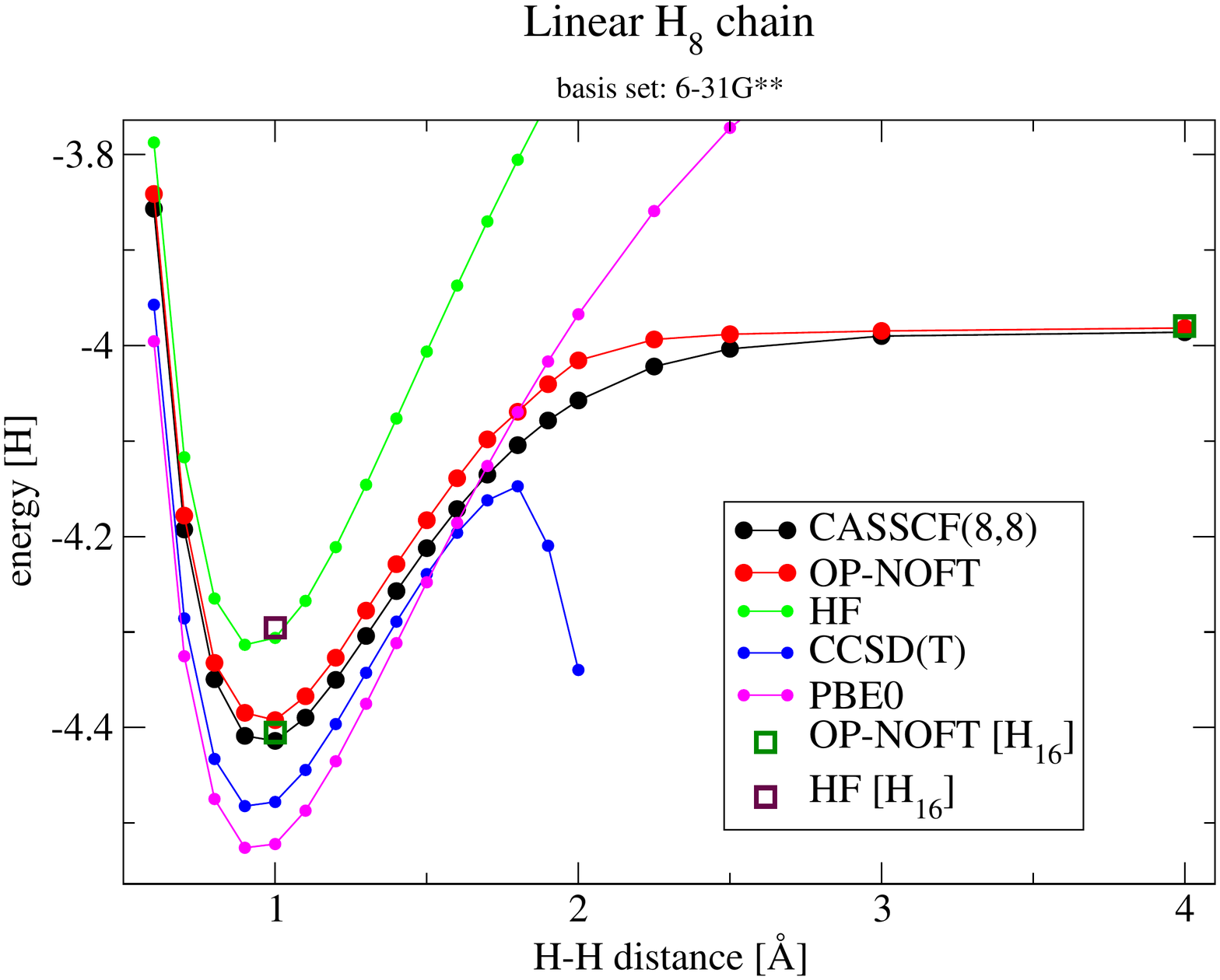}
  \caption{Symmetric dissociation curve of a linear H$_{8}$ chain.
    The squares indicate one half of the energy of a H$_{16}$ chain
    (black square: HF energy; green squares: OP-NOFT energy).
\label{fig1}}
\end{center}
\end{figure}

We consider linear H chains first. These are relatively simple systems
whose energy surfaces present a serious challenge for single reference
methods. Fig.~\ref{fig1} shows the dissociation energy curve of H$_8$
obtained with different methods. OP-NOFT-0 provides a consistent
description of the energy close to and everywhere above the CASSCF
reference. The breakdown of CCSD(T) at large separations is caused by
the single-reference character of this method.  The deviation of
OP-NOFT-0 from CASSCF should be attributed mainly to the restriction
to the $A=0$ sector, a conclusion supported by the
seniority-restricted CI calculations of
Ref.~\cite{Bytautas2011}. Close comparison with those calculations is
not entirely straightforward, as Ref.~\cite{Bytautas2011} used the
slightly smaller 6-31G basis and a fixed, symmetric or broken
symmetry, molecular orbital (MO) basis, whereas we used NOs that were
determined self-consistently.  Note that our results are better than
the CI results with $A=0,2$ and symmetric MO's. OP-NOFT-0 describes
the dissociation limit correctly.

\begin{figure}
\begin{center}
  \includegraphics[width= \columnwidth]{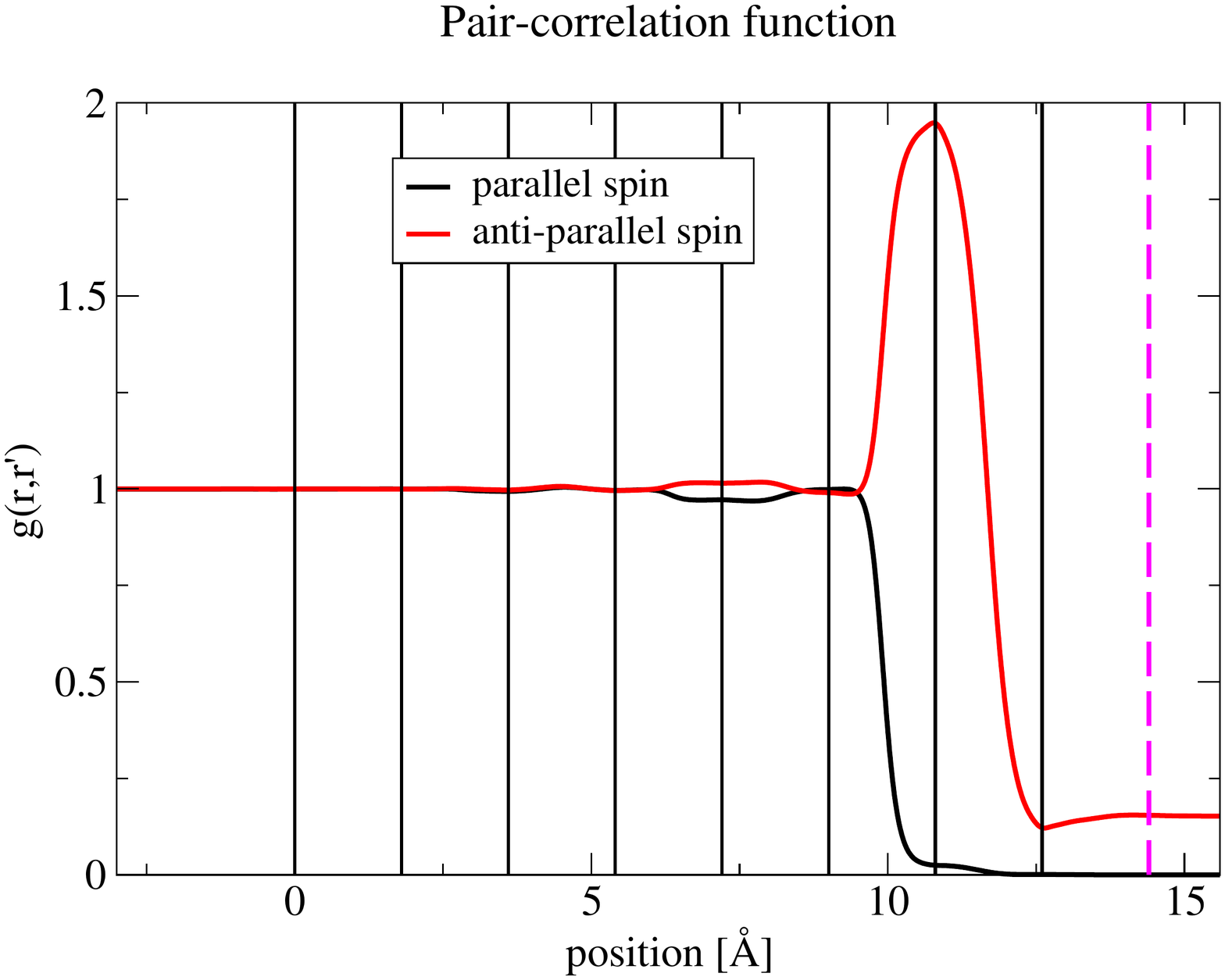}
  \caption{Electronic pair-correlation function along the H$_8$ axis
    when one electron is placed at the position of the vertical dashed
    line on the molecular axis. The vertical black lines show the atom
    positions. The H-H distance is 1.8 \AA.
\label{fig2}}
\end{center}
\end{figure}

OP-NOFT-0 provides not only the ground-state energy but also the 1-
and 2-DM. The former displays the entanglement due to correlation
through variation of the occupation numbers and the Von Neumann
entanglement entropy with interatomic separation shown in Fig.~S4 in
Section S5 of the SI. The increase of entanglement entropy with
separation signals a dramatic increase of correlation corresponding to
multi-reference character. The 2-DM gives access to electron pair
correlations.  Fig.~\ref{fig2} depicts the pair-correlation function
when one electron is placed outside the molecule's right end.  An
asymmetric exchange-correlation hole associated with the charging of
the end atom can clearly be seen. The alternating bonding/antibonding
character of the links between adjacent atoms is also manifest.  This
reflects the instability of the open-bounded chain toward a dimerizing
distortion and is evident in the pair correlations when an electron
resides respectively in the mid-bond or in the mid-antibond, as shown
in Fig.~S5, left and right panels respectively, in Section S6 of the
SI.

To test the dependence of the $\xi$ approximation~\rref{xi_DOCI} on
electron number, we studied the symmetric dissociation of
H$_{16}$~\footnote{We do not give the CASSCF energies in this case, as
  the dimension of the active subspace would make these calculations
  very expensive.}.  Results for the energy of H$_{16}$ divided by 2
are shown as squares in Fig.~\ref{fig1}.  OP-NOFT-0 works equally well
for this longer chain.  The total energy at dissociation is twice that
of H$_{8}$, and the slightly increased binding energy per atom at
equilibrium arises from an increase in the correlation energy, as
expected from more effective screening in the larger system.

\begin{figure}
\begin{center}
  \includegraphics[width=\columnwidth]{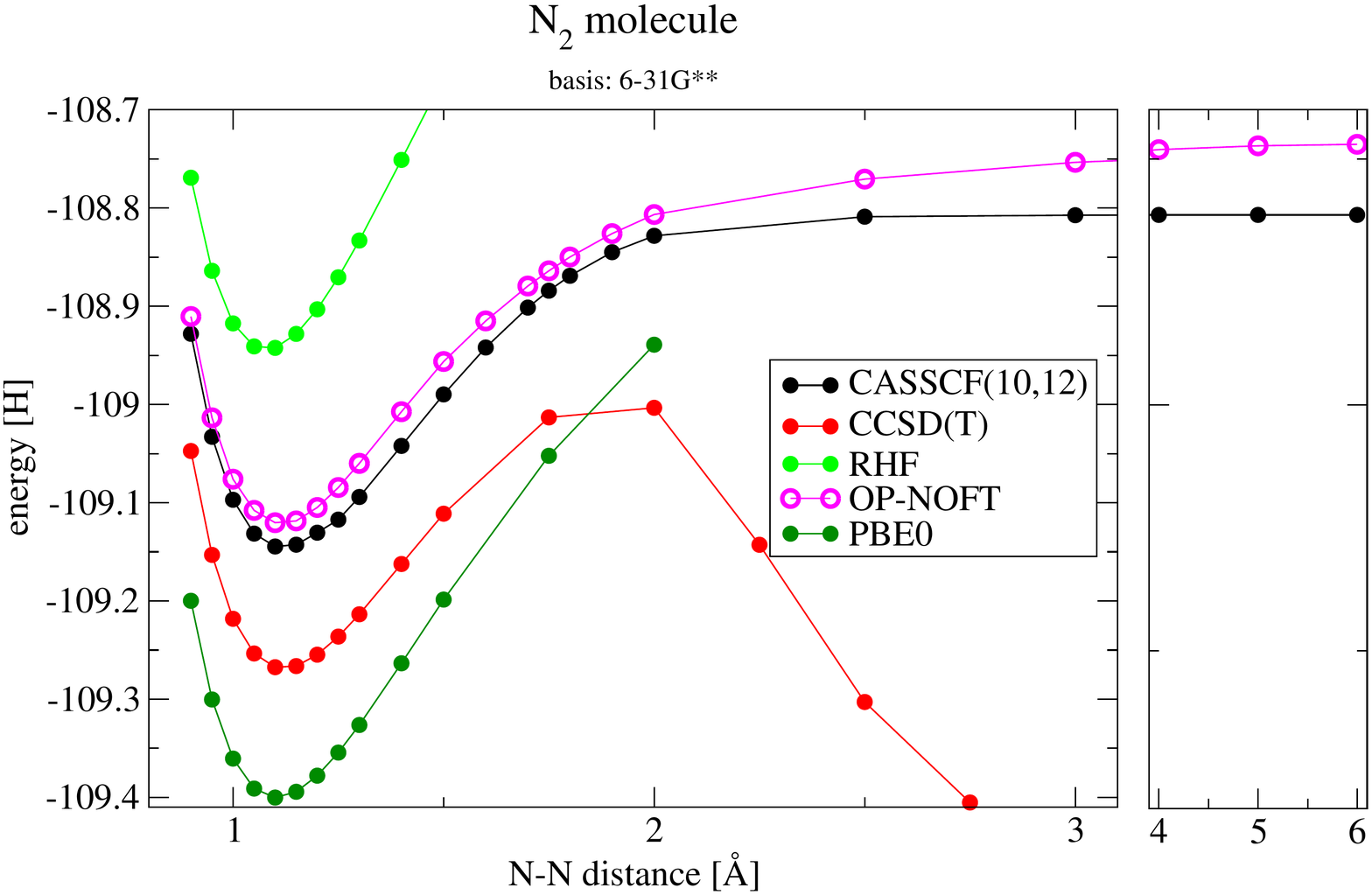}
  \caption{Dissociation curve of the N$_2$ molecule.
\label{fig3}}
\end{center}
\end{figure}

The N$_2$ molecule is a severe test for correlated electronic
structure methods because of its triple-bond. Our results are compared
to other methods in Fig.~\ref{fig3}. The OP-NOFT-0 curve is above the
CASSCF reference at all interatomic separations and deviates little
from it until $~2$\AA, beyond which the deviation increases with
separation until it stabilizes at $~5$\AA. OP-NOFT-0 correctly
dissociates the molecule into two non-interacting fragments, but
cannot capture correlations among the 3 electrons with unpaired spin
in each isolated atom.  It captures only intra-shell correlations
among electrons of opposite spin, consistent with seniority restricted
CI calculations in Ref.~\cite{Bytautas2011}.

It is interesting to note that in all the systems studied, the
positivity condition $(2,2)$ was found to be sufficient at near
equilibrium up to intermediate separations dominated by dynamic
correlation. Moreover, only in the case of H$_8$, H$_{16}$ and N$_2$ at
large 
separations did inclusion of the $(2,3)$ positivity condition turn out
to be essential to enforce numerically the variational character of
the ground-state solution.


\section{Discussion}

We have introduced a new method for correlated electronic-structure
calculations, OP-NOFT, that scales algebraically. Its DOCI-like
simplification, OP-NOFT-0 scales favorably with system size, with HF
energy-minimization scaling. The variational character of the energies
calculated via OP-NOFT-0 supports the accuracy of the $\xi$
approximation~\rref{xi_DOCI} and of limiting the positivity
conditions to \rref{bound1}--\rref{bound2}.  OP-NOFT-0 is restricted
to the $A=0$ sector of 
the Hilbert space. While providing an accurate description of
single-bond breaking and achieving a considerable improvement over
single-reference methods in all cases studied, it overestimates the
dissociation energy of a triple-bond, missing the correlations between
same-spin electrons in the open-shell fragments. That error should be
eliminated by including the contribution of the $A=2,4$ sectors as
shown in Ref.~\cite{Bytautas2011}.  Including those sectors would
require use of the full theory and its functional, \rref{E-od}. While
4-state NSO OPs and 4-index integrals would be required, the theory
would still scale polynomially.

It will be straightforward to add the computation of interatomic
forces to the OP-NOFT-0 energy-minimization methodology, making
possible the use of the theory for structural optimization and
ab-initio molecular dynamics~\cite{Car1985}.

From the practical point of view, minimization of the functional 
is significantly more laborious than minimization of the HF or the DFT
functional because considerably more minimization steps are needed to
minimize the functional \rref{energy}. We attribute this difficulty to the need
to include in \rref{energy} occupation numbers that are sufficiently small. In
damped dynamics minimization the forces acting on the corresponding NOs
are thus very weak compared to the forces acting on the NOs with
occupation numbers close to 1, slowing down considerably the entire
procedure. This difficulty is common to all NO-based methods including
those based on the 1-DM. Solving it is essential to making OP-NOFT
methods widely applicable in practice.

\begin{acknowledgments}
The authors acknowledge illuminating discussions with Paul Ayers and
Kieron Burke.  Refs.~\cite{Weinhold1967} and~\cite{Weinhold1967a} were
brought to the authors' attention by Paul Ayers.  The authors further
wish to thank J.~E.~Moussa for important comments.
M.H.C. and R.C. acknowledge support from the DOE under grant
DE-FG02-05ER46201.
\end{acknowledgments}





\bibliography{library}

\begin{thebibliography}{10}

\bibitem{Hohenberg1964}
Hohenberg P, Kohn W
\newblock (1964) {Inhomogeneous Electron Gas}.
\newblock \emph{Physical Review} 136:B864--B871.

\bibitem{Kohn1965}
Kohn W, Sham LJ
\newblock (1965) {Self-Consistent Equations Including Exchange and Correlation
  Effects}.
\newblock \emph{Physical Review} 140:A1133--A1138.

\bibitem{Gilbert1975}
Gilbert TL
\newblock (1975) {Hohenberg-Kohn theorem for nonlocal external potentials}.
\newblock \emph{Physical Review B} 12:2111--2120.

\bibitem{Gritsenko2005}
Gritsenko O, Pernal K, Baerends EJ
\newblock (2005) {An improved density matrix functional by physically motivated
  repulsive corrections.}
\newblock \emph{The Journal of chemical physics} 122:204102.

\bibitem{Lathiotakis2008}
Lathiotakis NN, Marques MAL
\newblock (2008) {Benchmark calculations for reduced density-matrix functional
  theory.}
\newblock \emph{The Journal of chemical physics} 128:184103.

\bibitem{Garrod1964}
Garrod C, Percus JK
\newblock (1964) {Reduction of the N-Particle Variational Problem}.
\newblock \emph{Journal of Mathematical Physics} 5:1756.

\bibitem{Coleman1963a}
Coleman AJ
\newblock (1963) {Structure of Fermion Density Matrices}.
\newblock \emph{Reviews of Modern Physics} 35:668--686.

\bibitem{Liu2007}
Liu YK, Christandl M, Verstraete F
\newblock (2007) {Quantum Computational Complexity of the N-Representability
  Problem: QMA Complete}.
\newblock \emph{Physical Review Letters} 98:110503.

\bibitem{Zhao2004}
Zhao Z, Braams BJ, Fukuda M, Overton ML, Percus JK
\newblock (2004) {The reduced density matrix method for electronic structure
  calculations and the role of three-index representability conditions.}
\newblock \emph{The Journal of chemical physics} 120:2095 -- 2104.

\bibitem{Mazziotti2012}
Mazziotti D
\newblock (2012) {Structure of Fermionic Density Matrices: Complete
  N-Representability Conditions}.
\newblock \emph{Physical Review Letters} 108:263002.

\bibitem{Nakata2001}
Nakata M, {et~al.}
\newblock (2001) {Variational calculations of fermion second-order reduced
  density matrices by semidefinite programming algorithm}.
\newblock \emph{The Journal of Chemical Physics} 114:8282.

\bibitem{Gidofalvi2008}
Gidofalvi G, Mazziotti DA
\newblock (2008) {Active-space two-electron reduced-density-matrix method:
  complete active-space calculations without diagonalization of the N-electron
  Hamiltonian.}
\newblock \emph{The Journal of chemical physics} 129:134108.

\bibitem{Weinhold1967}
Weinhold F, {Bright Wilson} E
\newblock (1967) {Reduced Density Matrices of Atoms and Molecules. I. The 2
  Matrix of Double-Occupancy, Configuration-Interaction Wavefunctions for
  Singlet States}.
\newblock \emph{The Journal of Chemical Physics} 46:2752.

\bibitem{Mazziotti2012a}
Mazziotti D
\newblock (2012) {Significant conditions for the two-electron reduced density
  matrix from the constructive solution of N representability}.
\newblock \emph{Physical Review A} 85:062507.

\bibitem{Ayers2007}
Ayers PW, Davidson ER
\newblock (2007) {Linear Inequalities for Diagonal Elements of Density
  Matrices}.
\newblock \emph{Advances in Chemical Physics} 134:443 -- 483.

\bibitem{Weinhold1967a}
Weinhold F, {Bright Wilson} E
\newblock (1967) {Reduced Density Matrices of Atoms and Molecules. II. On the
  N-Representability Problem}.
\newblock \emph{The Journal of Chemical Physics} 47:2298.

\bibitem{Davidson1969}
Davidson ER
\newblock (1969) {Linear Inequalities for Density Matrices}.
\newblock \emph{Journal of Mathematical Physics} 10:725.

\bibitem{Mazziotti2012b}
Mazziotti DA
\newblock (2012) {Two-electron reduced density matrix as the basic variable in
  many-electron quantum chemistry and physics.}
\newblock \emph{Chemical reviews} 112:244--62.

\bibitem{Bytautas2011}
Bytautas L, Henderson TM, Jim\'{e}nez-Hoyos CA, Ellis JK, Scuseria GE
\newblock (2011) {Seniority and orbital symmetry as tools for establishing a
  full configuration interaction hierarchy.}
\newblock \emph{The Journal of chemical physics} 135:044119.

\bibitem{Car1985}
Car R, Parrinello M
\newblock (1985) {Unified Approach for Molecular Dynamics and
  Density-Functional Theory}.
\newblock \emph{Physical Review Letters} 55:2471--2474.

\bibitem{Sheng2013}
Sheng XW, Mentel ÅM, Gritsenko OV, Baerends EJ
\newblock (2013) {A natural orbital analysis of the long range behavior of
  chemical bonding and van der Waals interaction in singlet H2: The issue of
  zero natural orbital occupation numbers}.
\newblock \emph{The Journal of Chemical Physics} 138:164105.

\bibitem{Perdew1996}
Perdew JP, Burke K, Ernzerhof M
\newblock (1996) {Generalized Gradient Approximation Made Simple}.
\newblock \emph{Physical Review Letters} 77:3865--3868.

\bibitem{Adamo1999}
Adamo C, Barone V
\newblock (1999) {Toward reliable density functional methods without adjustable
  parameters: The PBE0 model}.
\newblock \emph{The Journal of Chemical Physics} 110:6158.

\bibitem{Lowdin1956}
L\"{o}wdin PO, Shull H
\newblock (1956) {Natural Orbitals in the Quantum Theory of Two-Electron
  Systems}.
\newblock \emph{Physical Review} 101:1730--1739.

\end{thebibliography}

\end{article}








\end{document}